\documentclass[pr,11pt, reprint]{revtex4-1}

\usepackage{amsmath}    
\usepackage{graphicx}   
\usepackage{verbatim}   
\usepackage{color}      
\usepackage{subfigure}  
\usepackage{hyperref}   
\usepackage{upgreek}

\begin{document}

\title{Two-tone spectroscopy of a SQUID metamaterial in the nonlinear regime}
\author{E. I. Kiselev$^{1,2}$, A. S. Averkin$^3$, M. V. Fistul$^{3,4,5}$, V. P. Koshelets$^{6}$, A. V. Ustinov$^{2,3,5}$}
\affiliation{$^1$Institut f\"ur Theorie der Kondensierten Materie, Karlsruhe Institute of Technology, D-76131 Karlsruhe, Germany\\
$^{2}$Physikalisches Institut, Karlsruhe Institute of Technology, D-76131 Karlsruhe, Germany \\
$^{3}$National University of Science and Technology "MiSIS", 119049 Moscow, Russia\\
$^{4}$Center for Theoretical Physics of Complex Systems, Institute of Basic Science (IBS), Daejeon 34126, Republic of Korea\\
$^{5}$Russian Quantum Center, Skolkovo, Moscow 143025, Russia\\
$^{6}$Kotel'nikov Institute of Radio Engineering and Electronics, Moscow 125009, Russia}

\date{\today}

\begin{abstract}
Compact microwave resonantors made of superconducting rings containing Josephson junctions
(SQUIDs) are attractive candidates for building frequency tunable metamaterials with low losses
and pronounced nonlinear properties. We explore the nonlinearity of a SQUID metamaterial by
performing a two-tone resonant spectroscopy. The small-amplitude response of the metamaterial under
strong driving by a microwave pump tone is investigated experimentally and theoretically.
The transmission coefficient $S_{21}$ of a weak probe signal is measured in the presence of the pump tone.
Increasing the power of the pump, we observe pronounced oscillations of the SQUID's 
resonance frequency $f_{\textrm{res}}$. The shape of these oscillations varies significantly with
the frequency of the pump tone $f_{\textrm{dr}}$. The response to the probe signal displays
instabilities and sidebands. A state with strong second harmonic generation is observed.
We provide a theoretical analysis of these observations, which is in good agreement with 
the experimental results.
\end{abstract}

\maketitle

\section{Introduction}

Superconducting metamaterials have a number of unique properties that are difficult or impossible to
achieve in any other media \cite{JungUstinovAnlage}. A generic element of such a metamaterial, 
its elementary meta-atom, is a superconducting ring split by a Josephson junction. Each
meta-atom thus forms a Superconducting Quantum Interference Device (SQUID), the most relevant circuit element
of superconducting electronics. A Josephson junction acts as a nonlinear inductor,
the inductance depending on the superconducting current flowing through it. 
SQUIDs are used in a variety of applications including sensitive measurements of
magnetic fields and the detection of very small currents. In the microwave range, the unique properties of
SQUIDs include the tunability of their intrinsic resonance frequency \cite{Lazarides2007,LetterButz,1DMetaButz}, 
extremely low losses, and bi- and multistable response \cite{Likharev1986,Shnyrkov1980,Dmitrenko1982,Lazarides2013,JungFistul} 
allowing for a wireless switching between opaque and transparent states of the metamaterial \cite{Lazarides2013,JungFistul,AnlageTransparency}.

The nonlinear dynamics of Josephson junctions and SQUIDs plays an essential role in the development
of parametric amplifiers \cite{Castellanos,Yamamoto2008,Zorin,Zorin2017,Westig2018} and even evokes interest in fields as 
remote as biology \cite{NeuroSegall}. Parametric amplifiers are capable of amplifying one of the two quadratures of
a signal without adding extra noise. Their performance is
fundamentally limited only by quantum noise - an extraordinary property making
them relevant for applications in various fields ranging from astrophysics to the readout of qubits \cite{Devoret2016}. 
Driven beyond the parametric regime by a strong microwave driving, a
resonant Josephson Junction circuit exhibits a bifurcation and enters
the bi-stable regime with two stable dynamical states, which differ by both amplitude and phase.
This behavior has been used to build the so-called Josephson bifurcation amplifier \cite{Vijay}. 

Recently, a lot of attention
has been drawn to travelling wave parametric amplifiers (TWPAs), which can be realized with chains of Josephson junctions 
\cite{OBrien2014,Macklin2015} or SQUIDs \cite{Zorin,Zorin2017,Bell2015}. The SQUIDs of a TWPA chain have
to be strongly coupled, so that collective oscillations can be induced. In our case, there is no galvanic
connection between the SQUIDs. The coupling due to mutual inductance is very small, since the SQUIDs are spaced 
at large intervalls of $92\,\upmu\mathrm{m}$ - twice the width of a single SQUID. The distance to the central
conductor of the coplanar waveguide (CPW) is approximately $10\,\upmu\mathrm{m}$. Thus, the SQUIDs behave
as independent non-linear oscillators and are excited solely by the signal transmitted through the CPW. This
distinguishes our system from the Josephson junction and SQUID chains of Refs. \cite{Pop2010,Weissl2015,Krupko2018}, where
a crucial role is played by collective modes. 
We study the resonant response of the SQUID metamaterial
to a weak probe signal, while an additional strong driving (pump) signal pushes the metamaterial 
into a strongly nonlinear regime.
In contrast to studies of intermodulation
in SQUID metamaterials \cite{Anlage}, where two signals of close frequencies induce
sum and difference tones,
here we consider signals with remote frequencies and
are concerned with the strongly nonlinear behavior induced
by the pump. The frequency of the pump is fixed, while the weak signal
response is measured over a broad band of frequencies. We note that previously the authors of
Refs. \cite{Shnyrkov1980,Dmitrenko1982} presented
pioneering studies of the response of a single SQUID to a strong driving focusing
on the amplitude-frequency and amplitude-phase characteristics. The 
analysis given in Refs. \cite{Shnyrkov1980,Dmitrenko1982} resembles that of \cite{JungFistul}. 
Here, by extending the experiment and the theoretical framework to two applied signals and performing
a mathematical stability analysis of the solutions, we go beyond these works.
We find previously unobserved regions of instability in the response of the SQUIDs. Some of these instabilities
are associated with SQUID states that generate strong harmonics of the driving signal. Due to the large
signal amplitudes, these effects go beyond the scope of a description through Kerr coefficients \cite{Weissl2015,Krupko2018}.
The key to their understanding lies in the dynamics of strongly driven non-linear oscillators.

The paper is organized as follows: In section \ref{Main idea} we illustrate
the main physical ideas behind our study. Section \ref{Experiment} describes
the experimental setup, details of the measurement procedure, and presents our observations. In Section 
\ref{Theory} we provide the theoretical analysis of the nonlinear dynamics of an 
rf-SQUID in the presence of both the weak probe signal and the large amplitude driving tone. 
We start from revisiting a previously developed analytical approach to the nonlinear SQUID response \cite{JungFistul},
and then apply it's
formalism to the two-tone scenario. Section IV concludes the paper.

\section{The main idea}
\label{Main idea}

\begin{figure}[ht]
    \centering
    \includegraphics[width=.13\textwidth,angle=-90]{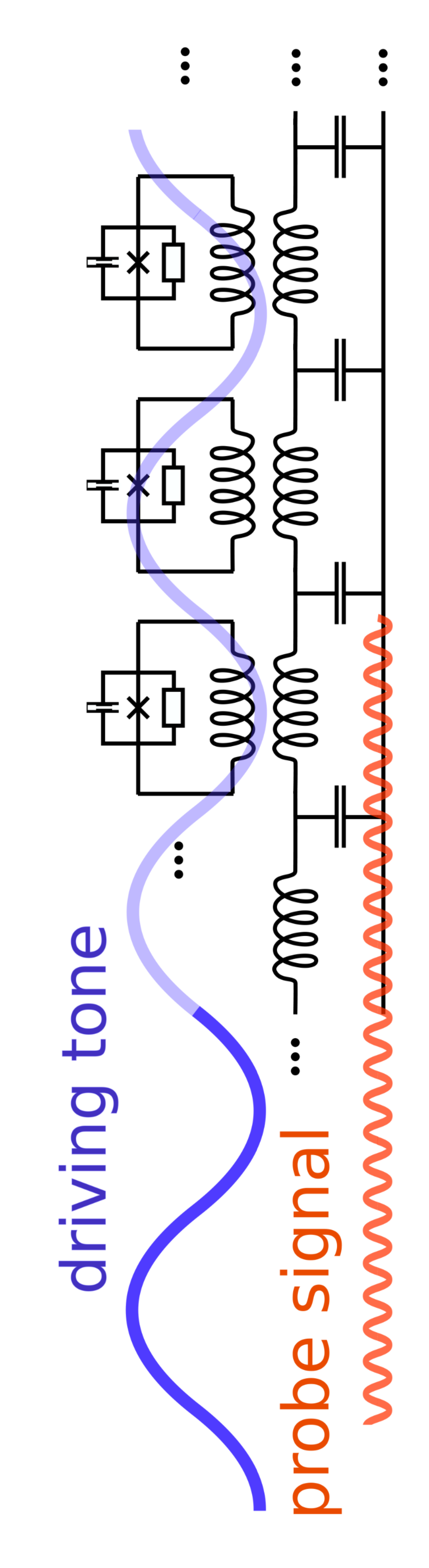}
    \caption{The main idea: the nonlinear response of the metamaterial
    to a large amplitude driving tone (blue) is investigated by a weak
    probe signal (orange). The response to the weak signal is linear,
    however it contains information about the large amplitude response
    to the driving tone.}
    \label{main_idea}
\end{figure}

The main idea behind our work is that the large amplitude, steady state response of a nonlinear
system to a strong driving can be investigated by a weak probe signal, which is superimposed on the 
driving (see Fig. \ref{main_idea}). The oscillatory response to the the weak probe signal is expected to be small, and hence 
governed by a linear equation. The parameters of this equation (most relevantly the resonance frequency) will sensitively
depend on the amplitude and frequency of the strong driving. Thus, by examining the weak signal
response, we gain information about the nonlinear steady state oscillations.

It is known that the response of (non-hysteretic) rf-SQUIDs - our meta-atoms - in a certain range of driving 
amplitudes and frequencies, has regions with multiple stable steady states, 
a feature known as multistability \cite{JungFistul}. 
We find that a distinct resonance dip in the response to the probe signal corresponds
to each of the stable states, allowing us to map the multistabilities.
Furthermore, unstable regions, in which no stable steady-state solutions exist, are seen as gaps within which no resonant
response to the weak probe signal is observed.

\section{Experimental Setup and Results}
\label{Experiment}

\subsection{Experimental Setup}

\begin{figure}[ht]
    \centering
    \includegraphics[width=.4\textwidth]{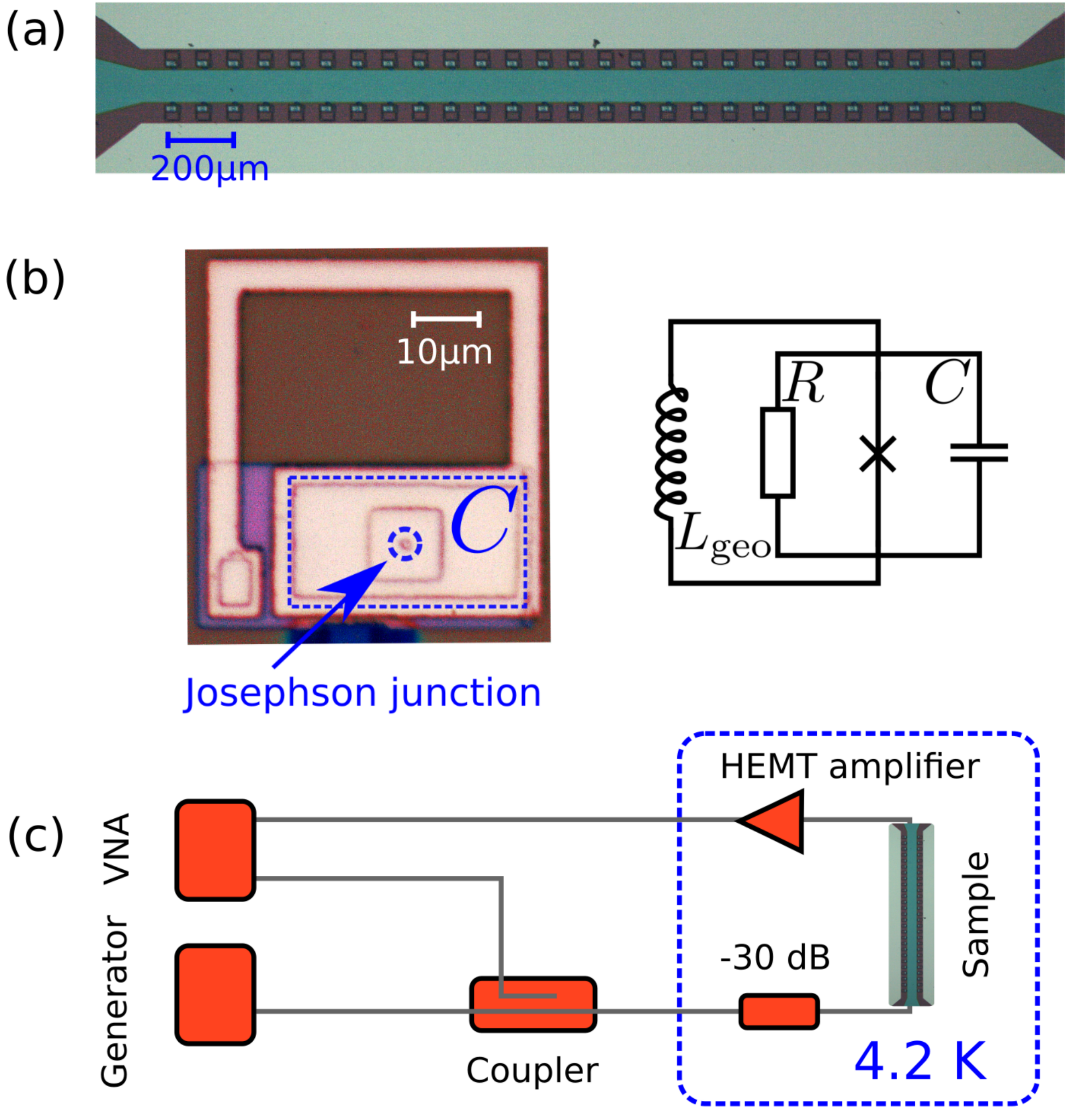}
    \caption{(a) An image of the rf-SQUID metamaterial showing 54 rf-SQUIDs coupled
    to a coplanar waveguide. (b) Left: Photo of a single rf-SQUID. Right:
    Equivalent circuit model of an rf-SQUID (RCSJ model \cite{Tinkham}). (c) The experimental setup consisting of
    a $^4$He cryostat at $4.2\ \textrm{K}$, a microwave generator (GEN) and vector
    network analyzer (VNA).}
    \label{ChipAndSetup}
\end{figure}

The experiments were carried out at a temperature of $4.2\textrm{ K}$ using a one dimensional
array of $54$ single junction SQUIDs (rf-SQUIDs). In this setting, each SQUID
constitutes a metamaterial atom embedded in a CPW. This is shown
in Fig. \ref{ChipAndSetup} (a). The SQUIDs
(see Fig. \ref{ChipAndSetup} (b)) were fabricated using the well established Nb/AlO$_x$/Nb
trilayer process. The SQUID parameters are the same as reported in previous publications and experiments \cite{1DMetaButz}. 
The value of the Josephson critical current is $I_{\textrm{c}}=1.8\ \mu\textrm{A}$, leading to a
zero field Josephson inductance of $L_{\textrm{j},0} = 82.5\ \textrm{pH}$. The geometric
loop inductance amounts
$L_{\textrm{geo}} = 183\ \textrm{pH}$ and the shunt capacitance is $C = 2.0\ \textrm{pF}$. These characteristics
result in a resonance frequency of approximately
$14.8\ \textrm{GHz}$ at zero magnetic field. The ratio of zero field Josephson
and geometric inductances is $\beta_{\textrm{L}}=2\pi I_\textrm{c}L_\textrm{geo}/\Phi_0\approx0.45$,
where $\Phi_0=h/(2e)$ is the magnetic flux quantum. The value $\beta_\textrm{L}<0$ indicates
that the SQUIDs are in the non-hysteretic regime, thus having each a single stable static state at any magnetic
flux.

A strong microwave pump signal generated by a
microwave source (the driving tone) was inserted into the CPW to carry out the two-tone spectroscopy of
the above described sample. In the presence of this driving
tone, we measured the transmission coefficient $S_{21}$ of a weak probe signal
propagating through the sample. A vector network analyzer (VNA)
(see, Fig. \ref{ChipAndSetup} (c)) was used for these measurements.
Typical frequencies for
the two signals were $1-10\ \textrm{GHz}$ and $14-20\ \textrm{GHz}$ for the driving tone
and $10-16\ \textrm{GHz}$ for the probe signal. No external magnetic field was applied.

\subsection{Observations}

\begin{figure}[ht]
    \centering
    \includegraphics[width=.45\textwidth]{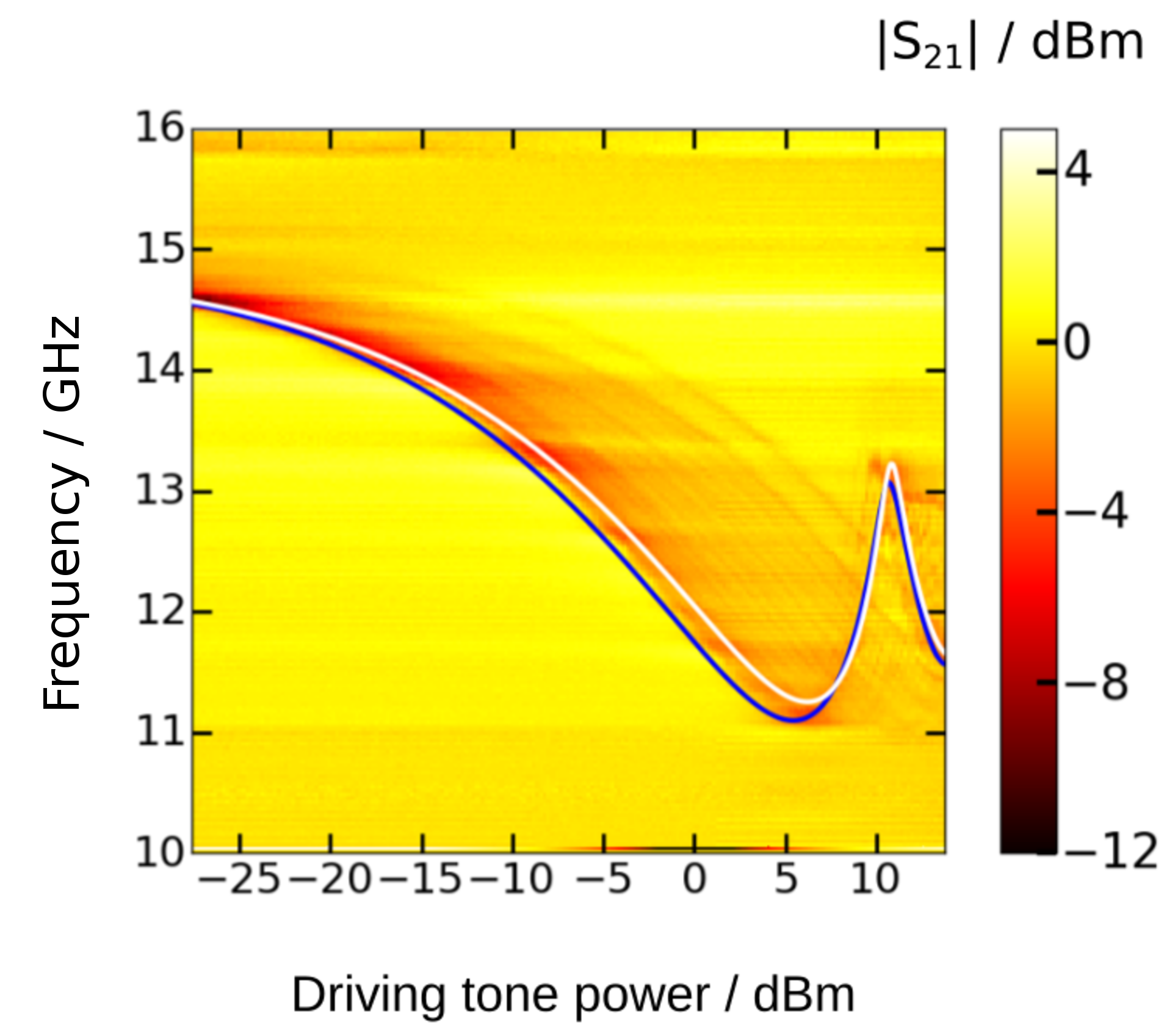}
    \caption{The transmission spectrum of the metamaterial at a driving tone frequency 
    $f_{\textrm{dr}}=15\textrm{ GHz}$. Driven into a strongly nonlinear regime by a large
    amplitude driving tone, the SQUID metamaterial shows a resonant
    response to the weak probe signal of the VNA.
    The resonance frequency is seen as a minimum in the transmission
    coefficient $S_{\textrm{21}}$. With increasing driving tone power
    the resonance frequency drops first, then shows an oscillatory behavior. The
    blue curve shows the theoretically calculated resonance
    frequency as given by the harmonic approximation of Eq.
    (\ref{MeasResFreq}), while the white curve shows the result obtained by solving 
    Eq. (\ref{ResCond}).}
    \label{Exp15GHz}
\end{figure}

\begin{figure}[ht]
    \centering
    \includegraphics[width=.45\textwidth]{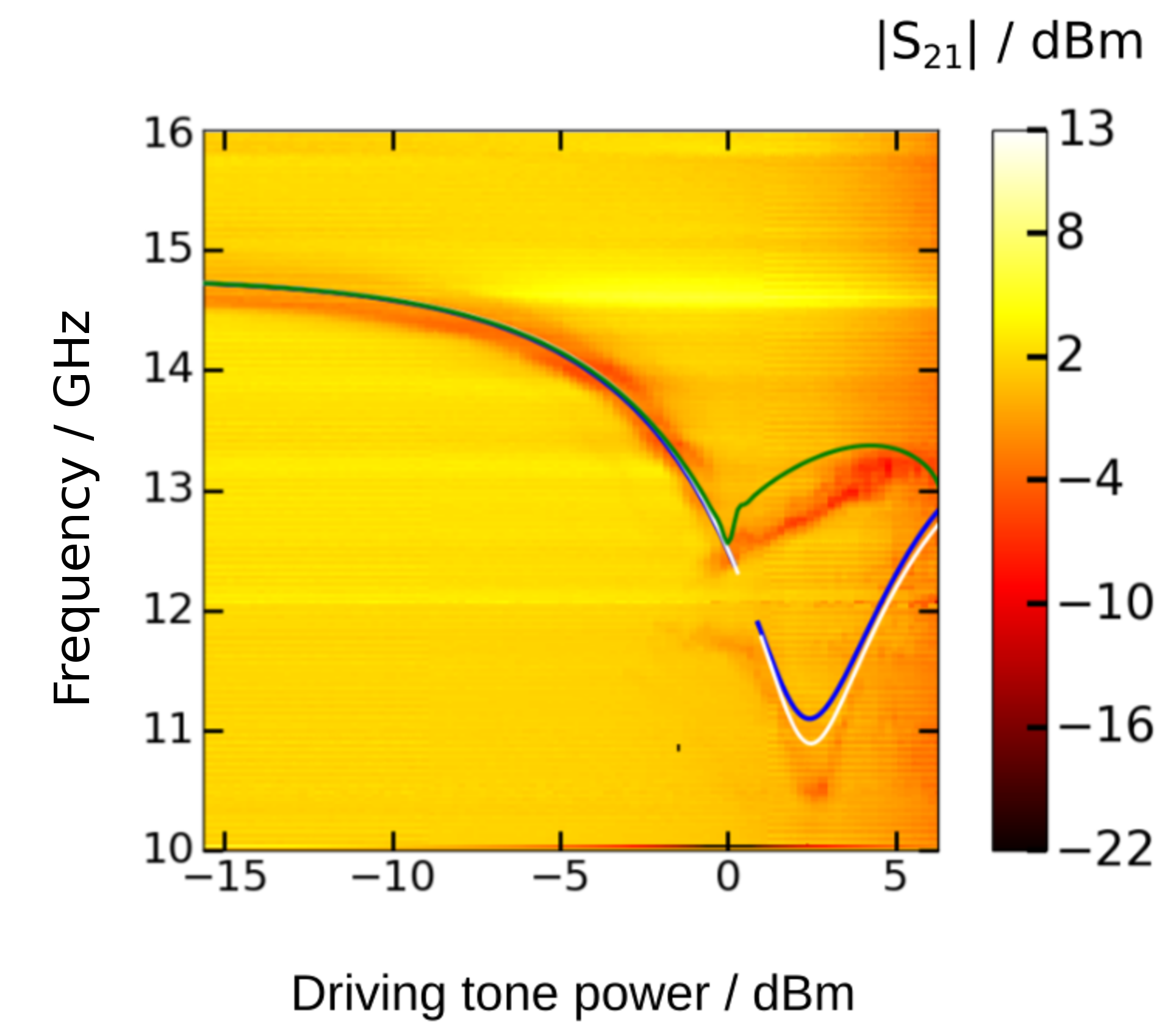}
    \caption{The transmission spectrum of the metamaterial at a driving tone frequency 
    $f_{\textrm{dr}}=6\textrm{ GHz}$. Apart from the typical oscillatory behavior of the resonance
    frequency, we observe a gap in the resonance curve around $12\ \textrm{GHz}$, at approximately
    $0\ \textrm{dBm}$ driving power. This gap indicates an unstable region in the response of the SQUID.
    In the vicinity of the gap,
    the resonance curve splits into two branches corresponding to two stable states that can
    be occupied by the SQUIDs - the response is bistable. The
    blue curve shows the resonance frequency as calculated using Eq.
    (\ref{MeasResFreq}) and is not plotted in the region of instability
    predicted by Eq. (\ref{m=2Instability}). The white curve shows the resonance frequency
    obtained from Eq.
    (\ref{ResCond}). In the
    bistable region, the green curve, which was calculated
    using Eq. (\ref{two-harmonics-resonance-curve}), depicts the resonance associated with a
    state that is characterized by a strong response at twice the
    driving frequency (see section \ref{Bistability in the response}).}
    \label{Exp6GHz}
\end{figure}

\begin{figure}[ht]
    \centering
    \includegraphics[width=.45\textwidth]{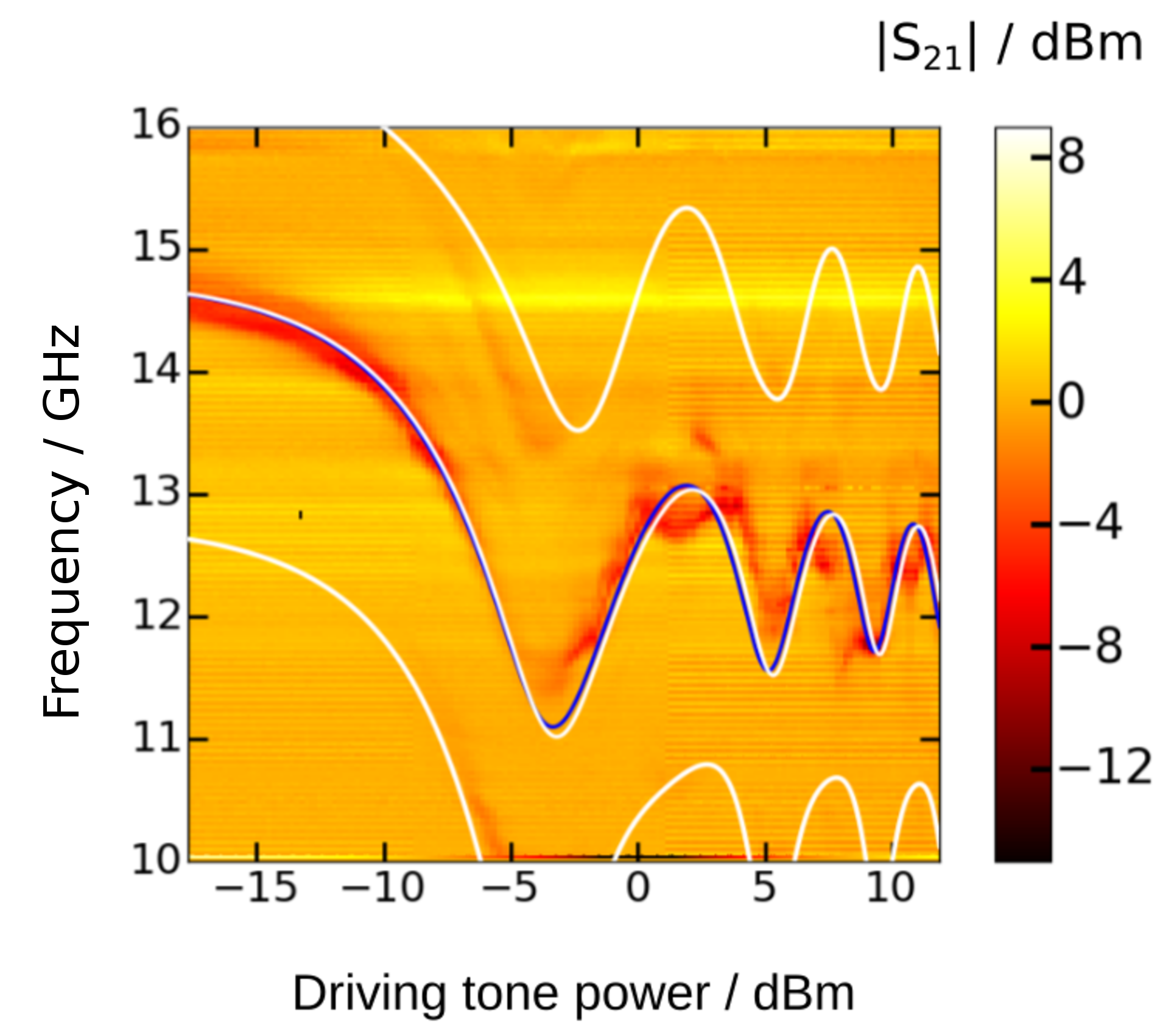}
    \caption{Transmission spectrum of the metamaterial at a driving tone frequency 
    $f_{\textrm{dr}}=1\textrm{ GHz}$. 
    Sideband resonances appear at
    $f_{\textrm{ms},0}\pm2\ \textrm{GHz}$, where $f_{\textrm{ms},0}$ {is} the
    resonance frequency at a given power of the driving tone.
    The blue curve shows the main resonance as given by Eq.
    (\ref{MeasResFreq}). The white curve shows the resonance frequencies obtained
    from Eq. (\ref{ResCond}). Eq. (\ref{ResCond}) includes the effect of sum and difference
    tones of the driving and probe signals. Including these sum and difference tones gives
    a natural explanation for the resonant sidebands.}
    \label{Exp1GHz}
\end{figure}

We observed that the frequency dependent transmission coefficient
$S_{21}$ of the weak probe signal shows, even in the presence of a strong pump
signal, a narrow resonant drop at the resonance
frequency $f_{\textrm{res}}$. The resonance frequency $f_{\textrm{res}}$ itself strongly
varies with the power of the externally applied pump tone. Typical
experimentally measured dependencies of $S_{21}(f_{\textrm{pr}}, P)$ on the
frequency $f_{\textrm{pr}}$ of the weak probe signal and the power $P$ of the
driving tone are presented in Figs. \ref{Exp15GHz}-\ref{Exp1GHz}, for driving tone frequencies of
$15\ \textrm{GHz}$, $6\ \textrm{GHz}$ and $1\ \textrm{GHz}$, respectively.

For all driving tone frequencies, we observed
a substantial decrease of the resonance frequency $f_{\textrm{res}}$ with increasing driving tone power.  At high powers, 
the resonance frequency $f_{\textrm{res}}$ showed an oscillatory behavior.

An additional feature seen in Fig. \ref{Exp15GHz} is the splitting of the
resonance curves of different SQUIDs. This splitting is due to the slightly
different strength of coupling of each single SQUID to the CPW. If the coupling
of a SQUID is comparatively weak, the effect of the driving tone on this
particular SQUID also will be smaller. Thus, its resonance curve will be shifted
to higher values of the driving tone power.

At lower frequencies of the driving tone ($f_{\textrm{dr}} \leq 7 \textrm{ GHz}$) we observe gaps 
in the probe signal's resonant response. Such a gap is seen in Fig. \ref{Exp6GHz} at $12\ \textrm{GHz}$
and a driving power of around $0\ \textrm{dBm}$. It indicates an
instability in the response of the strongly driven SQUIDs, as will be explained below.
In the vicinity of the gap the resonance curve splits.
This is due to the existence of two stable SQUID states in this range of parameters.
Both states are occupied by some of the 54 SQUIDs of our metamaterial, allowing us to 
image the bistability directly.

Lowering the driving frequency even further, we observe resonant sidebands
in the probe signal's transmission coefficient $S_{21}$ (see Fig. \ref{Exp1GHz}).
These sidebands are located above and below the main resonance curve,
at frequencies $f_{\textrm{res}}\pm 2 f_{\textrm{dr}}$, where $f_{\textrm{dr}}$ is the
frequency of the driving tone and $f_{\textrm{res}}$ is the resonance frequency
at a given power of the driving tone.

All above observations can be explained by an analysis of the nonlinear
rf-SQUID dynamics in the presence of a large amplitude microwave signal. The
oscillations of the resonance frequency with the power of a driving tone are obtained by
treating the response to the probe signal as that of a harmonic
oscillator with parameters depending on the power of the driving tone. To obtain the
resonance gaps and sidebands, we have to go beyond this harmonic model and turn to a more
detailed treatment of the generic SQUID nonlinearity.

\section{Theoretical Description}
\label{Theory}
\subsection{The nonlinear response of rf-SQUIDs to a harmonic driving}
\label{Nonlinear response of rf-SQUIDs to harmonic driving}

In order to explain the effects reported in the previous section we assume
that the interaction between adjacent rf-SQUIDs is negligible, and that therefore
the electrodynamic response of the superconducting metamaterial is determined by
the response of individual SQUIDs. The dynamics of an rf-SQUID is
governed by a simple
cirquit model: the resistively and capacitively shunted Josephson junction
(RCSJ) model \cite{Tinkham} in parallel with the geometric inductance of the
SQUID ring $L_{\textrm{geo}}$ (see Fig. \ref{ChipAndSetup} (b)).
In this model the time-dependent Josephson phase $\varphi(t)$ satisfies
the nonlinear equation:
\begin{equation}
    \varphi_{\textrm{ext}} = \varphi + L_{\textrm{geo}}C\ddot{\varphi} +
    \frac{L_{\textrm{geo}}}{R}\dot{\varphi}
    +\beta_L \sin{\varphi}\ ,
    \label{SQUID_eq}
\end{equation}
where  $\varphi_{\textrm{ext}}$ consists of both, an externally applied driving
tone and a probe signal, $\beta_L=L_{\textrm{geo}}/L_{\textrm{j}}$,
$L_{\textrm{j}}=\Phi_0/2\pi I_{\textrm{c}}$ with $I_c$
being the critical current and $R$ the normal resistance of the Josephson junction.

We start by considering the nonlinear dynamics of an rf-SQUID in the presence
of a strong driving tone $\varphi_{\textrm{ext}}=\varphi_{\textrm{dr}}\sin(\omega_{\textrm{dr}}t)$.
Approximate solutions to Eq. (\ref{SQUID_eq}) are obtained by exploiting the fact
that the response of an rf-SQUID to a driving tone of frequency
$\omega_{\textrm{dr}}$ will mostly be dominated by harmonic oscillations at
the same frequency $\omega_{\textrm{dr}}$.
This approach has already been successfully applied to a study of dynamic metastable
states that can be excited in an rf-SQUID metamaterial \cite{JungFistul} and
uses the following ansatz
\begin{align}
    \nonumber
    \varphi_{\textrm{ext}} &= \varphi_{\textrm{dr}}\sin{\omega_{\textrm{dr}} t}
    \\
    \varphi(t) &= \varphi_{\textrm{a}}\sin{(\omega_{\textrm{dr}} t + \delta)}\ .
    \label{ansatz}
\end{align}
The validity of this ansatz has been analyzed in the methods section of Ref. \cite{JungFistul}. It was found
that the ratio between the first and third harmonics is numerically small, which justifies the monochromatic
approach of Eq. (\ref{ansatz}).
Notice that $\varphi_{\textrm{dr}} \propto \sqrt{P}$ for the driving tone amplitude
holds, where $P$ is the power of the driving tone.
Inserting the ansatz of Eq. (\ref{ansatz}) into Eq. (\ref{SQUID_eq}) the nonlinear term of
Eq. (\ref{SQUID_eq}), is expanded into a Fourier series with the help of a Jacobi-Anger identity \cite{AbraSteg}
\begin{align}
    \nonumber
    &\sin{(\varphi_{\textrm{a}}\sin{(\omega_{\textrm{dr}} t + \delta)})} =
    \\
    &2\sum^\infty_{n=0}J_{2n+1}(\varphi_{\textrm{a}})\sin{\big((2n+1)
    (\omega_{\textrm{dr}} t + \delta)\big)}\ ,
    \label{NonlinearFourier}
\end{align}
where $J_n(x)$ are Bessel functions of $n$th-order. 
All terms of this expansion except for $n=0$ are neglected according to
our assumptions and Eq.
(\ref{SQUID_eq}) reads
\begin{align}
    \nonumber
    \varphi_{\textrm{dr}}\sin{\omega_{\textrm{dr}} t} &=
    \Bigg(1-\frac{\omega_{\textrm{dr}}^2}{\omega^2_{\textrm{geo}}}\Bigg)\varphi_{\textrm{a}}
    \sin{(\omega_{\textrm{dr}} t+\delta)}
    \\
    \nonumber
    & + \frac{\omega_{\textrm{dr}}}{\omega_{\textrm{c}}}\varphi_{\textrm{a}}\cos{(\omega_{\textrm{dr}} t+\delta)}
    \\
    &+ 2\beta_L J_1(\varphi_{\textrm{a}})\sin{(\omega_{\textrm{dr}} t+\delta)}\ ,
    \label{Inter1}
\end{align}
where $\omega_{\textrm{c}} = R/L_{\textrm{geo}}$ and the geometrical frequency is given by
$\omega_{\textrm{geo}}=(L_{\textrm{geo}}C)^{-1/2}$.
Rewriting Eq. (\ref{Inter1}) in terms of $\sin{(\omega_{\textrm{dr}}t)}$ and
$\cos{(\omega_{\textrm{dr}}t)}$ and comparing the coefficients of these functions we obtain an
equation for the amplitude of the driving tone $\varphi_{\textrm{dr}}$ in terms of
the amplitude of the Josephson phase response $\varphi_{\textrm{a}}$:
\begin{equation}
    \varphi_{\textrm{dr}} =
    \sqrt{\Bigg[\Bigg(1-\frac{\omega_{\textrm{dr}}^2}{\omega_{\textrm{geo}^2}}\Bigg)
    \varphi_{\textrm{a}} + 2\beta_LJ_1(\varphi_{\textrm{a}})\Bigg]^2
    + \frac{\omega_{\textrm{dr}}^2}{\omega^2_{\textrm{c}}}\varphi^2_{\textrm{a}}}\ .
    \label{AmpEq}
\end{equation}
For the SQUIDs used here the resistance $R$ in is of the order of $10\,\textrm{k}\Omega$
and hence $\omega_{\textrm{c}}\approx 50\cdot 10^{12}/\textrm{s}$. Thus, to a
good approximation, the last term in Eq. (\ref{AmpEq}) can be neglected.

\begin{figure}[h]
    \centering
    \includegraphics[width=.35\textwidth]{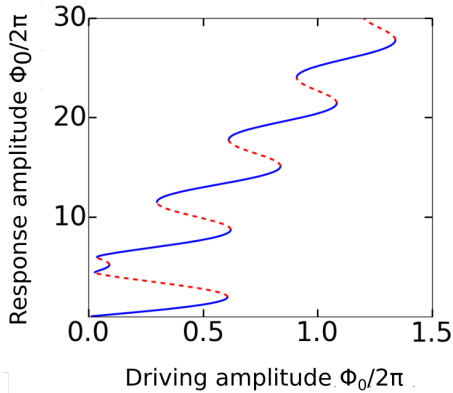}
    \caption{The response amplitude of a SQUID to a harmonic external
    driving at the frequency $\omega_{\textrm{dr}}\approx0.998\cdot\omega_{\textrm{geo}}$
    as given by formula (\ref{AmpEq}). The stable regions of 
    $\varphi_{\textrm{a}}(\varphi_{\textrm{ea}})$ are plotted in blue, whereas in the
    red dotted areas the solution of Eq. (\ref{AmpEq}) is unstable.
    For some values of the driving amplitude, response at different amplitudes
    is possible - a feature generally referred to as multistability. 
    The parameters $\beta_L=0.45$ and $\omega_{\textrm{geo}}=14.8\textrm{ GHz}$ were chosen.}
    \label{StabilityInstability}
\end{figure}

From Eq. (\ref{AmpEq}) one can see that if the SQUID is driven at a frequency near
the geometric resonance, i.e. $\omega_{\textrm{dr}}\approx\omega_{\textrm{geo}}$, we have
$\varphi_{\textrm{dr}} \approx 2\beta_L|J_1(\varphi_{\textrm{a}})|$.
Since $J_1(\varphi_{\textrm{a}})$ is an oscillating function, the inverse function
$\varphi_{\textrm{a}}(\varphi_{\textrm{dr}})$ is multi-valued. This phenomenon is referred to as
\textit{multistability} and was experimentally observed by Jung et al.
\cite{JungFistul}, as well as by Refs. \cite{Shnyrkov1980,Dmitrenko1982}, however,
without putting an emphasis on it. The response amplitude of a SQUID as a
function of the driving amplitude $\varphi_{\textrm{a}}(\varphi_{\textrm{dr}})$ for
$\omega_{\textrm{dr}}\approx0.998\cdot\omega_{\textrm{geo}}$ is shown in Fig. \ref{StabilityInstability}.
Some parts of the curve $\varphi_{\textrm{a}}(\varphi_{\textrm{dr}})$ are unstable in the sense, that any
small perturbation to the amplitude $\varphi_{\textrm{a}}$ will cause a change of
the state of the SQUID. These parts are marked with red dotted lines in Fig. \ref{StabilityInstability}.

To investigate the stability of the solutions of Eq.
(\ref{AmpEq}), we followed a standard procedure \cite{BogolyubovMitropolsky},
where the time evolution of a small perturbation to the phase and amplitude
of the SQUID response as given by Eq. (\ref{ansatz})
is  studied. The response is written
$\varphi_{\textrm{a}}\sin{(\omega_{\textrm{dr}} t + \delta)} =
x\sin{\omega_{\textrm{dr}} t} + y\cos{\omega_{\textrm{dr}} t}$
with $\varphi_{\textrm{a}} = \sqrt{x^2+y^2}$, $\delta = \arctan(y/x)$ and the
perturbations $\lambda_i(t)$ {are} added: $x\rightarrow x + \lambda_1(t)$,
$y\rightarrow y + \lambda_2(t)$, $\lambda_i(t)\ll 1$. Linear differential
equations are then derived for
$\lambda_i(t)$ from Eq. (\ref{SQUID_eq}) assuming the validity of Eq. (\ref{AmpEq}).
In unstable regions the perturbations $\lambda_i(t)$ show an exponential growth, while stability is
assured, if only oscillatory solutions exist for $\lambda_i(t)$. The results agree
with the general expectations \cite{BogolyubovMitropolsky,JungFistul,Anlage}, that those sections
of $\varphi_{\textrm{a}}(\varphi_{\textrm{dr}})$ are unstable, where the curve's slope is negative.

\subsection{Two-tone resonant spectroscopy of rf-SQUIDs}

In this subsection, we analyze the resonant response of the superconducting metamaterial to
both, the weak probe signal and the strong driving tone. First, we
will treat the response within a harmonic approximation, i.e. we will approximate the probe
signal response as that of a harmonic oscillator with parameters 
depending on the driving tone amplitude. This approximation is
sufficient to explain the oscillations of the SQUID's resonance frequency
$f_\textrm{res}$ with the increasing power of the
driving tone. Next, the sidebands and unstable regions seen in the experiments will be
described within a more detailed analysis.

\subsubsection{Harmonic approxiamtion}
\label{Harmonic approxiamtion}

The two signals applied to the SQUID are the driving tone
$\varphi_{\textrm{dr}}\sin{(\omega_{\textrm{dr}} t)}$ and the probe signal of the VNA
$\varphi_{\textrm{pr,ext}}$. The response is assumed to be of the form
$\varphi = \varphi_{\textrm{a}}\sin{(\omega_{\textrm{dr}} t +\delta)}+\varphi_{\textrm{pr}}$.
Eq. (\ref{SQUID_eq}) then takes the following form:
\begin{widetext}
\begin{align}
    \nonumber
    \varphi_{\textrm{dr}}\sin(\omega_{\textrm{dr}} t) + \varphi_{\textrm{pr,ext}} =&
    \varphi_{\textrm{a}}\sin(\omega_{\textrm{dr}} t + \delta)
    + \varphi_{\textrm{pr}}
    -
    \frac{\omega_{\textrm{dr}}^2}{\omega^2_{\textrm{geo}}}
    \varphi_{\textrm{a}}\sin(\omega_{\textrm{dr}} t + \delta)
    +
    \omega^{-2}_{\textrm{geo}}\ddot{\varphi}_{\textrm{pr}}
    \\
    &+
    \frac{\omega_{\textrm{dr}}}{\omega_{\textrm{c}}}\varphi_{\textrm{a}}\cos(\omega_{\textrm{dr}} t + \delta)
    +
    \omega_{\textrm{c}}^{-1}\dot{\varphi}_{\textrm{pr}}
    +
    \beta_L \sin{(\varphi_{\textrm{a}}\sin(\omega_{\textrm{dr}} t + \delta) + \varphi_{\textrm{pr}})}\ .
    \label{SQUID_eqSIGN}
\end{align}
\end{widetext}
For a weak probe signal, the response $\varphi_{\textrm{pr}}$ will be small.
Expanding up to the first order in $\varphi_{\textrm{pr}}$, the nonlinear term of
(\ref{SQUID_eqSIGN}) becomes
\begin{equation}
    \sin{(\varphi_{\textrm{a}}\sin(\omega_{\textrm{dr}} t + \delta))}
    + \cos{(\varphi_{\textrm{a}}\sin(\omega_{\textrm{dr}} t + \delta))}\cdot\varphi_{\textrm{pr}}\ .
    \label{Inter2}
\end{equation}
At this point a key assumption enters the calculation: since the VNA signal is
very weak and $|\varphi_{\textrm{pr}}|\ll\varphi_{\textrm{a}}$, its presence does not
significantly alter the state of the SQUID. It is hence assumed that the 
Eq. (\ref{Inter1}) is still
valid. Substituting Eqs. (\ref{Inter1}) and (\ref{Inter2}) into Eq. (\ref{SQUID_eqSIGN})
we obtain an effective equation for the response $\varphi_{\textrm{pr}}$ to the VNA
signal:
\begin{align}
    \nonumber
    \varphi_{\textrm{pr,ext}} &=
    \varphi_{\textrm{pr}}
    +
    \omega_{\textrm{geo}}^{-2}\ddot{\varphi}_{\textrm{pr}}
    +
    \omega_{\textrm{c}}^{-1}\dot{\varphi}_{\textrm{pr}}
    \\
    &+
    \beta_L
    \cos{(\varphi_{\textrm{a}}\sin(\omega_{\textrm{dr}} t + \delta))}\cdot\varphi_{\textrm{pr}}\ .
    \label{TimeTerms}
\end{align}
This equation with a time dependent, periodic coefficient is known as Hill's equation
\cite{Arscott}. Solutions of the homogeneous part of this
equation can be unstable in the sense that their amplitude exponentially grows in time
\cite{Arscott,LL1}. These instabilities, however, only occur for specific values
of $\omega_{\textrm{dr}}$, $\varphi_{\textrm{a}}$ and $\omega_{\textrm{c}}$ and will be studied
in next subsection.
Here, solely the special solution to (\ref{TimeTerms})
is of interest, and only the time average of $\cos{(\varphi_{\textrm{a}}\sin(\omega
t + \delta))}$ over one period, i.e. the zeroth Fourier mode, will be
taken into account. We use the Jacobi-Anger identity \cite{AbraSteg}
\begin{align}
    \nonumber
    &\cos{(\varphi_{\textrm{a}}\sin{(\omega_{\textrm{dr}} t + \delta)})} =
    \\
    &J_{0}(\varphi_{\textrm{a}})
    +2\sum^\infty_{n=1}J_{2n}(\varphi_{\textrm{a}})\cos{\big(2n
    (\omega_{\textrm{dr}} t + \delta)\big)}\ .
    \label{NonlinearFourier2}
\end{align}
And approximate the Eq. (\ref{TimeTerms}) by
\begin{align}
    \varphi_{\textrm{pr,ext}}
    =
    [1+\beta_LJ_0(\varphi_{\textrm{a}})]\varphi_{\textrm{pr}}
    +
    \omega_{\textrm{c}}^{-1}\dot{\varphi}_{\textrm{pr}}
    +
    \omega_{\textrm{geo}}^{-2}\ddot{\varphi}_{\textrm{pr}}\ .
    \label{LinearizedSignalEq}
\end{align}
The frequency of the resonant drop of the probe signal's transmission coefficient
$S_{21}$ then is
\begin{equation}
    f_{\textrm{res}} =
    \frac{\omega_{\textrm{geo}}}{2\pi}
    \sqrt{[1+\beta_LJ_0(\varphi_{\textrm{a}})]
    -\frac{\omega^2_{\textrm{geo}}}{2\omega^2_{\textrm{c}}}}\ .
    \label{MeasResFreq}
\end{equation}
This result gives the shift of the resonance frequency as a function of the amplitude
of the driving response $\varphi_{\textrm{a}}$ (see Eq. (\ref{AmpEq})). In the
experimental data, however, the transmission coefficient $S_{21}$ is plotted against
the power of
the driving tone, $P$. We thus introduce a coupling constant $C$ by means of
\begin{equation}
    \varphi_{\textrm{dr}} = C\sqrt{P}\ .
\end{equation}
When fitting Eq.
(\ref{MeasResFreq}) to experimental data, $C$ is the only free
parameter. It has to be obtained  independently for every driving tone
frequency, since the transmission through microwave cables and devices is
frequency dependent.

In Figs. \ref{Exp15GHz}-\ref{Exp1GHz} the SQUID
resonance frequency predicted by Eq. (\ref{MeasResFreq}) is plotted blue. Unsurprisingly,
the analysis presented so far describes the experiment best
at high driving tone frequencies (Fig. \ref{Exp15GHz}), since in this regime the
time dependent terms in the expansion of Eq. (\ref{NonlinearFourier2}) are expected to
average out. However, even at low driving tone frequencies the Eq. (\ref{MeasResFreq})
is in a good accord with experimental results (Figs. \ref{Exp6GHz} and \ref{Exp1GHz}).

\subsubsection{Parametric instabilities of the SQUID response}

At certain values of the driving frequency $\omega_{dr}$ and the driving amplitude
$\varphi_{\textrm{dr}}$ the solution to the
homogeneous part of equation (\ref{TimeTerms}) becomes unstable \cite{Arscott}. To describe this
phenomenon called parametric instability, or parametric
resonance \cite{LL1}, we take into account the
$n=1$
term in the expansion of Eq. (\ref{NonlinearFourier2}) which oscillates with the frequency
of $2\omega_{\textrm{dr}}$. The effective probe signal equation
(\ref{TimeTerms}) then becomes
\begin{align}
    \nonumber
    \varphi_{\textrm{pr,ext}}
    &=
    [1+\beta_LJ_0(\varphi_{\textrm{a}})+J_2(\varphi_{\textrm{a}})
    \cos{(2\omega_{\textrm{dr}} t)}]\varphi_{\textrm{pr}}
    \\
    &+
    \omega_{\textrm{c}}^{-1}\dot{\varphi}_{\textrm{pr}}
    +
    \omega_{\textrm{geo}}^{-2}\ddot{\varphi}_{\textrm{pr}}\ .
    \label{MyMathieu}
\end{align}
It is well known \cite{Arscott,LL1}, that the parametrically unstable regions of
such a Mathieu-type equation occur at driving frequencies of
\begin{equation}
    \omega_{\textrm{dr}} = \frac{\omega_{\textrm{res}}}{m} \pm\Delta\omega_{\pm}\ ,
    \label{BraggCondition}
\end{equation}
where $m$ is an integer number, and $\Delta\omega_{\pm}$ is a small frequency
interval whose value depends on
the driving tone amplitude $\varphi_{\textrm{a}}$. Applying the stability criterion
of Ref. \cite{LL1} to Eq. (\ref{MyMathieu}), estimations on
$\Delta\omega_{\pm}$ are obtained.
For $m=1$ we have
\begin{equation}
    \Delta\omega_{\pm} =
    \Bigg|\frac{\beta_LJ_2(\phi_{\textrm{a}})\omega_{\textrm{res}}}
    {2(1+\beta_LJ_0(\phi_{\textrm{a}}))}\Bigg|\ .
    \label{m=1Instability}
\end{equation}
The $m=1$ instability occurs when $\omega \approx
\omega_{\textrm{res}}\approx \omega_{\textrm{geo}}$ holds for the driving frequency. It is therefore not
surprising that the intervals given by Eq. (\ref{m=1Instability}) coincide with the instability regions already
discussed in the Subsection \ref{Nonlinear response of rf-SQUIDs to harmonic driving}.

For the $m=2$ instability, however, we observe a different behavoir. In this case
the interval of instability is not symmetric:
\begin{align}
    \nonumber
    \Delta\omega_{+} &=
    \frac{1}{24}\Bigg|\frac{\beta_LJ_2(\phi_{\textrm{a}})\omega_{\textrm{res}}}
    {2(1+\beta_LJ_0(\phi_{\textrm{a}}))}\Bigg|^2
    \\
    \Delta\omega_{-} &=
    \frac{5}{24}\Bigg|\frac{\beta_LJ_2(\phi_{\textrm{a}})\omega_{\textrm{res}}}
    {2(1+\beta_LJ_0(\phi_{\textrm{a}}))}\Bigg|^2\ .
    \label{m=2Instability}
\end{align}
Figure \ref{Exp6GHz} shows the
experimental data obtained at a driving tone frequency of $6\ \textrm{GHz}$. The
$m=2$ instability is clearly seen as a gap in the resonance curve around $12\ \textrm{GHz}$ at a driving power of
approximately $0\ \textrm{dBm}$ . The gap in the blue curve which shows the resonance frequency as given by Eq.
(\ref{MeasResFreq}) indicates the unstable region predicted by Eq. (\ref{m=2Instability}).
Furthermore, higher order instabilities were
observed for $m=3$ and $m=4$ at driving frequencies of $4\ \textrm{GHz}$ and $3\ \textrm{GHz}$
respectively. While reaching an instability in the $m=1$ case just indicates a
jump to another stable state predicted by Eq. (\ref{AmpEq}), in the higher order
cases (\ref{AmpEq}) does not predict any multistabilities.
Thus the $m=2,3,...$
instabilities suggest that the SQUID response in the region of these instabilities is more
complicated than assumed in deriving Eq. (\ref{AmpEq}). 
In fact, we demonstrate below that 
the resonant drop seen in Fig. \ref{Exp6GHz} above the gap and at powers
above $0\ \textrm{dBm}$ indicates that a fraction of the SQUIDs occupies a state where
harmonics with frequencies $\omega_{\textrm{dr}}$ and $2\omega_{\textrm{dr}}$ are
approximately equally strong. Such a state indeed goes beyond the 
approximations leading to (\ref{AmpEq}).

\subsubsection{A new, symmetry breaking bistability in the SQUID response}
\label{Bistability in the response}

To explain the bistability observed at a driving frequency of 6 GHz 
(see Fig. \ref{Exp6GHz}) it is necessary to go beyond the approximate ansatz of Eq.
(\ref{ansatz}). The bistability appears when, while increasing the driving amplitude,
the resonance frequency $f_{\textrm{res}}$ is tuned to twice the driving
frequency $f_{\textrm{dr}}$. At the same time the frequency of the second harmonic
of the response with a frequency of $2f_{\textrm{dr}}$, too, becomes approximately equal 
to the resonance frequency. The amplitude of the second harmonic can therefore
be expected to grow in this regime.
In fact, numerics show that the resonant drop in Fig. \ref{Exp6GHz}, which 
is not described by Eqs. (\ref{MeasResFreq})
or (\ref{ResCond}), results from a SQUID state where the amplitudes 
$\varphi_{\textrm{a}1}$, $\varphi_{\textrm{a}2}$,
of the first and second harmonics with frequencies 
$\omega_{\textrm{dr}}$ and $2\omega_{\textrm{dr}}$ have the same order
of magnitude. We therefore use the Ansatz
\begin{align}
    \varphi(t) &= \varphi_{\textrm{a}1}\sin{(\omega_{\textrm{dr}} t + \delta_1)}
    +\varphi_{\textrm{a}2}\sin{(2\omega_{\textrm{dr}} t + \delta_2)}
    \label{ansatz_2}
\end{align}
to describe this state.
The frequency of the resonant drop is obtained analogously to
(\ref{MeasResFreq}) and given by
\begin{align}
    \nonumber
    f_{\textrm{res}} = &\frac{\omega_{\textrm{geo}}}{2\pi}\Big(1 + \beta_L
    J_0(\varphi_{\textrm{a}1})J_0(\varphi_{\textrm{a}2}) 
    \\
    &+ 2\beta_L\sum_{n=1}^{\infty}
    J_{4n}(\varphi_{\textrm{a}1})J_{2n}(\varphi_{\textrm{a}2})\cos{(2n(2\delta_1-\delta_2))}\Big)^{1/2}\ .
    \label{two-harmonics-resonance-curve}
\end{align}
The amplitudes
$\varphi_{\textrm{a}1}$ and $\varphi_{\textrm{a}2}$ are obtained
numerically as Fourier coefficients of the first and second
harmonics of the solution to Eq. (\ref{SQUID_eq}). For the numerical analysis, we used 
the standard fourth order Runge-Kutta method, and generated a solution $\varphi(t)$ of Eq. (\ref{SQUID_eq}). 
To find the amplitudes
$\varphi_{a1}$ and $\varphi_{a2}$ we projected the solution $\varphi(t)$ onto the Fourier modes, e.g.:
\begin{align}
    \nonumber
    \varphi^2_{a2} =& \left(\frac{1}{n}\int_0^{nT}dt\,\sin{(2\omega t)}\varphi(t)\right)^2\\
    &+\left(\frac{1}{n}\int_0^{nT}dt\,\cos{(2\omega t)}\varphi(t)\right)^2,
\end{align}
where $T=2\pi/\omega$ and $n$ is the number of periods over which the solution was averaged.
The green solid line in Fig. (\ref{Exp6GHz}) shows the resonance as described by Eq.
(\ref{two-harmonics-resonance-curve}) in comparison with experimental
results. 

The fact that a strong second harmonic can be generated by the SQUIDs without a dc magnetic field being applied
is rather remarkable. Such a state dynamically breaks the $\varphi\rightarrow-\varphi$ centrosymmetry of Eq. (\ref{SQUID_eq}).
Dynamical symmetry breaking is a phenomenon known from the theory of non-linear oscillators \cite{Olson1991,Miles1989}. 
However, it has not yet been observed in SQUID dynamics to our best knowledge. In fact, the
observation of this bistability demonstrates the advantages of the
two-tone spectroscopy over frequency-amplitude and frequency-phase measurements, in which
it cannot be imaged with such ease. We expect that other irregularities in the weak signal response,
like e.g. the ones seen in Fig. \ref{Exp1GHz}, can similarly be explained in terms of states with a strong
response at harmonics of the driving tone.

\subsubsection{Sideband resonances}

Figure \ref{Exp1GHz} shows the response to the probe signal at a low driving 
tone frequency $\omega_{\textrm{dr}}\ll\omega_{\textrm{pr}}$. We observe
sideband resonances approximately 2 GHz above and below the main resonance.
These sidebands appear due to the term $J_2(\varphi_{\textrm{a}})\cos{(2\omega_{\textrm{dr}} t)}$
in the probe signal equation (\ref{MyMathieu}).
The term mixes the oscillation at $\omega_{\textrm{pr}}$ with modes at frequencies
separated from it by a multiple of $2\omega_{\textrm{dr}}$.
To calculate the positions of the sidebands we therefore use the ansatz
\begin{align}
    \nonumber
    \varphi (t) &= a \sin{(\omega_{\textrm{pr}} t)} + b_1\sin{(\omega_{\textrm{pr}}
    t+2\omega_{\textrm{dr} } t)}
    \\
    &+ b_2\sin{(\omega_{\textrm{pr}} t-2\omega_{\textrm{dr} } t)}\
    \label{ThreeModeApproximation}
\end{align}
and $\varphi_{\textrm{pr,ext}} = \varphi_{\textrm{pr}}\sin{(\omega_{\textrm{pr}}t)}$.
To describe the sidebands nearest to the main resonance, modes with
frequencies further away from $\omega_{\textrm{pr}}$,
i.e. $\omega_{\textrm{pr}}\pm 4\omega_{\textrm{dr} }$,
$\omega_{\textrm{pr}}\pm 6\omega_{\textrm{dr} }$, and so on can be neglected.
Since the dissipation is rather small, one can neglect phase shifts $\delta$
in the above ansatz.

We substitute (\ref{ThreeModeApproximation}) into (\ref{MyMathieu}) and solve for
$a(\varphi_{\textrm{pr}})$. With dissipation being neglected,
resonances occur at those frequencies $\omega_{\textrm{pr}}$
at which $a(\varphi_{\textrm{pr}})\propto\varphi_{\textrm{pr}}$ diverges, or, equivalently,
\begin{equation}
    \varphi_{\textrm{pr}}(a) = 0\ .
    \label{ResCond}
\end{equation}
Solving Eq. (\ref{ResCond}) involves finding the roots of a sixth order polynomial
in $\omega_{\textrm{pr}}$, which is done numerically. For most values of the driving
tone frequency $\omega_{\textrm{dr}}$ and power $P$, the equation (\ref{ResCond}) has
three positive roots, corresponding to the main resonant response
and \emph{two sidebands}. The resonance and sideband curves predicted by Eq.
(\ref{ResCond}) for the driving tone frequency of $1\ \textrm{GHz}$ are plotted
in Fig.
\ref{Exp1GHz} (the white solid line) and are in good agreement with the experiment. 
For higher driving tone frequencies (Fig. \ref{Exp6GHz}) the sidebands are 
outside of the experimentally accessible frequency range.
The main resonance curves given by Eq. (\ref{ResCond}) (white solid line) and Eq.
(\ref{MeasResFreq}) (blue solid line) agree with each other rather well.
In unstable regions, the roots found solving Eq. (\ref{ResCond}) are complex, and
hence no resonances can be observed. The region of parametric instability
determined in this way is consistent with the region determined by the conditions
of Eqs. (\ref{m=1Instability}), (\ref{m=2Instability}).

\section{Conclusions}

We presented the results of a two-tone spectroscopy of an rf-SQUID metamaterial consisting
of 54 single SQUIDs placed in a transmission line. The power of the pump (driving) tone
was typically much higher than the power of the probe signal used to measure 
the transmission spectrum. 

We observed that the
resonance frequency of the metamaterial, seen as a drop in the transmission
spectrum of the probe signal at frequencies between 10 and 15 GHz, shows a 
characteristic  oscillatory dependence on the power of a pump tone. 
The frequency of the pump tone was varied between 1 GHz and 20
GHz, changing the shape of the resonance curve. 

For pump tone frequencies below or in the region of  the resonance frequencies, i.e. 
for our parameters $f_{\textrm{dr}}<14.5$ GHz,
we observed the gaps in the resonance curves of the transmission spectrum.
A bistability in the response was directly observed at an intermediate range of 
the pump frequency (~6 GHz).
At a low pump frequencies $f_{\textrm{dr}}$ sidebands located approximately $2f_{\textrm{dr}}$  
above and below the main resonance curve were found.

Most of our observations are well described theoretically by an approach based on an
approximate analytical model for the response of a SQUID to strong external
driving \cite{JungFistul}. Whereas the shape of the resonance curves naturally followed
from an extension of the model to two tones of different amplitudes, the gaps
could be explained as parametric instabilities of the SQUID response. 
The observed bistability is shown to appear due to the existence of a symmetry breaking
state with strong second harmonic generation.

We believe that the decribed effects have multiple applications. The instability
induced gaps can be used to create metamaterials with tunable transparancy. The symmetry breaking
second harmonic generation can be employed in frequency doublers. Last but not least, the remarkably 
clear small signal response at even the highest pump powers shows that a study
of parametric amplification in the highly non-linear regime could be a worthwhile endeavor.

To summarize, at high driving power levels SQUID metamaterials exhibit a rich spectrum of 
features tunable by the power and the frequency of the pump tone. The non-linear two tone 
spectroscopy is a powerfull method to study these features, which can be applied
to other systems. The high degree of nonlinearity and 
the unique multistable behavior make superconducting circuits an exciting playground for studies of
tunable metamaterials.

\begin{acknowledgments}
We thank B. Jeevanesan and A. Zorin for helpful discussions and L.
Filippenko for the fabrication of the experimental circuits. The support of this work by the
Deutsche Forschungsgemeinschaft (DFG) (Grant No. US 18/15-1) is gratefully acknowledged. 
We also acknowledge partial support by the Ministry of Education and Science of Russian Federation 
in the framework of Increase Competitiveness Program of the NUST MISiS (Grant No. K2-2017-081). 
Work concerning cirquit fabrication was supported by the RSF (Project No. 19-19-00618).
\end{acknowledgments}

\end{document}